\documentclass[12pt,aps,prf,onecolumn,groupedaddress,noshowpacs]{revtex4-2}

\usepackage{bm} 
\usepackage{color}
\usepackage{amsmath}
\usepackage{amssymb}
\usepackage{mathrsfs}
\usepackage{graphicx} 
\usepackage{lastpage}
\usepackage{epstopdf}
\usepackage[normalem]{ulem}

\begin{document}

\newcommand{\blue}[1]{{\color{blue}#1}}
\newcommand{\red}[1]{{\color{red}#1}}
\newcommand{\magenta}[1]{{\color{magenta}#1}}

\title{Comment on ``Consistent hydrodynamics of ferrofluids"}

\author{Mark I. Shliomis}
\email{shliomis@bgu.ac.il, mshliomis@gmail.com}

\affiliation{Department of Mechanical Engineering,
Ben-Gurion University of the Negev, Beer-Sheva 84105, Israel}

\date{\today}

\maketitle

In a recent paper [1], Fang presented ``a generic and consistent hydrodynamic theory for ferrofluids" which ``is expected to be a cornerstone of ferrofluid hydrodynamics." Here is how he formulates the main flaw of existing theory: ``Unfortunately, in the ferrofluid community all previous studies fail to discriminate between the solvent vorticity ${\bm\Omega}_{sol}$ and the suspension vorticity ${\bm\Omega}$." By his definition, the solvent vortices are  unobserved quantities. To observe (measure) them, it would be necessary to extract magnetic particles from the ferrofluid, and replace them with holes, i.e., turn ferrofluid into a kind of Swiss cheese with holes instead of magnetic grains. Therefore, ``${\bm\Omega}_{sol}$, the mesoscopic vorticity due to solvent flow, is not a genuine macroscopic variable and should be eliminated," the author writes.

But how to do that? ``How to correctly express ${\bm\Omega}_{sol}$ in terms of macroscopically quantities characterizing the suspension?" And Fang found a ``simple but important relation" that relates ${\bm\Omega}_{sol}$ to ${\bm\Omega}$ and the mean angular velocity of particles ${\bm\omega}$ by a linear relationship
\begin{eqnarray}
{\bm\Omega}=(1-\phi){\bm\Omega}_{sol}+\phi\kern 1pt{\bm\omega}\,;
\end{eqnarray}
here $\phi=nV_p$ and $n$ are the volume fraction and the number density of magnetic particles, $V_p=\pi d^3/6$.

Unfortunately, the relation (1) contradicts to the angular momentum conservation law and therefore is false. Indeed, when two physical objects (A and B), rotating with frequencies ${\bm\omega}_A$ and ${\bm\omega}_B$, merge into one (C), then, according to the law of conservation of angular momentum, the contributions of ${\bm\omega}_A$ and ${\bm\omega}_B$ to ${\bm\omega}_C$ are determined by the ratio of the moments of inertia $I_A/I_B$, and not by the relative volume fraction $\phi$, as it is done in Eq.~(1).

To derive the correct relation, one should use the conservation law for the angular momentum:
\begin{eqnarray}
I_{FF}{\bm\Omega}=I_{FF}{\bm\Omega}_0+I_{MP}({\bm\omega}-
{\bm\Omega}_0),
\end{eqnarray}
where $I_{FF}$ is the moment of inertia of the ferrofluid, and $I_{MP}$ is the sum of the intrinsic moments of inertia of the magnetic particles contained in it; ${\bm\Omega}$ and ${\bm\Omega}_0$ are ferrofluid vortices in and without an external magnetic field. In the absence of a field, the particles rotate together with the fluid, ${\bm\omega}={\bm\Omega}_0\,$, so ferrofluid rotates as a whole: ${\bm\Omega}={\bm\Omega}_0\,$. Constant magnetic field always hampers particle's rotation, ${\omega<\Omega_0}$, while a rotating magnetic field can provide the inverse inequality.

Dividing Eq.~(3) by $I_{FF}$, we arrive at the relation
\begin{eqnarray}
{\bm\Omega}=\Big(1-\frac{I_{MP}}{I_{FF}}\Big){\bm\Omega}_0+
\frac{I_{MP}}{I_{FF}}\kern 1pt{\bm\omega},
\end{eqnarray}
that differs from (1) in two ways: by replacing $\phi$ by $I_{MP}/I_{FF}$ and by replacing the unobserved ${\bm\Omega}_{sol}$ to ${\bm\Omega}_0$. Let us evaluate these replacements.

The moment of inertia of one magnetic particle is equal $\rho_pV_p(d^2/10)$, their number in a cylinder of radius $R$ and length $S$ is $\pi nR^2S$, so their total moment of inertia is
\begin{eqnarray*}
I_{MP}=(\pi/10)\rho_p\kern 1pt\phi\kern 1pt d^2R^2S.
 \end{eqnarray*}
The moment of inertia of the ferrofluid filling the cylinder (below $\rho_s$ and $\rho_p$ are the densities of the solvent and particles) is equal to
\begin{eqnarray*}
I_{FF}=(\pi/2)\big(\rho_s(1-\phi)+\rho_p\phi\big)R^4S\,,
\end{eqnarray*}
and the ratio of the above moments of inertia is
\begin{eqnarray}
\frac{I_{MP}}{I_{FF}}=C\phi\Big(\frac{d}{R}\Big)^2,~~~{\rm where}~~~
C=\frac{\rho_p/5}{\rho_p\phi+\rho_s(1-\phi)}\,.
\end{eqnarray}
Substituting this relation in Eq.~(3), we finally find
\begin{eqnarray}
{\bm\Omega}=\Big(1-C\phi\frac{d^2}{R^2}\Big){\bm\Omega}_0+
C\phi\frac{d^2}{R^2}{\bm\omega}.
\end{eqnarray}
For particles of size $d\simeq 20\,{\rm{nm}}$ and vessel radius $R\simeq 2\,{\rm{cm}}$ we have
$(d/R)^2\simeq 10^{-12}\kern 1pt$(!). Since $C$ is always of the order of unity, Eq.~(5) reduces to ${\bm\Omega}={\bm\Omega}_0\,$; this equality holds with high accuracy for any values of $\phi$. Thus, the uniform rotation of the nanoparticles does not in itself affect the rotation rate of the ferrofluid; the reasons for this have been explained in \cite{Shliomis-21}. Nonetheless, the spin of particles may turn on indirect mechanisms of ferrofluid rotation \cite{Shliomis-21}, which are not covered in \cite{Fang-22}.

Using the erroneous initial assumption (1), Fang immediately arrived at the mistaken equation for the ferrofluid magnetization
\begin{eqnarray}
\frac{d{\bf M}}{dt}=\Big(\frac{1}{1-\phi}{\bm\Omega}+\frac{\phi}{1-
\phi}{\bm\omega}\Big)\times{\bf M}-
\frac{\bf M-\bf M_0}{\tau_M}\,,
\end{eqnarray}
and then to the mistaken expression \cite{notation} for the dimensionless rotational viscosity $\eta_R=[\eta(H)-\eta(0)]/\eta(0)$:
\begin{eqnarray}
\eta_R=\frac{3}{2}\phi\kern 1pt\frac{\alpha L^2(\alpha)}{(\alpha-L(\alpha))F(\alpha)}~~~~{\rm where}~~~~
F(\alpha)=1-\phi\kern 1pt\frac{\alpha L^2(\alpha)}{\alpha-L(\alpha)}\,;
\end{eqnarray}
here $L(\alpha)$ is the Langevin function. Fang compared Eq.~(7) with the result
\begin{eqnarray}
\eta_R=\frac{3}{2}\phi\kern 1pt\frac{\alpha L^2(\alpha)}{\alpha-L(\alpha)},
\end{eqnarray}
obtained in \cite{Shliomis-01} on the basis of the Martsenyuk-Raikher-Shliomis theory \cite{Martsenyuk et al.}. As can be seen, these equations differ from each other by the additional function $F(\alpha)$ in the denominator of Eq.~(7). The absence of this function in Eq.~(8),  Fang explains by the fact that ``Shliomis neglects the difference between the vorticity of the suspension and the vorticity of the solvent." This is true: as shown above, neither this difference nor the solvent vorticity itself exists.

Assumption (1) is the main, but not the only mistake made in \cite{Fang-22}. Fang is constantly confused about the rotational speeds of particles ${\bm\omega}$, fluid ${\bm\Omega}$, and magnetization ${\bm\omega}_M$. ``Should we identify ${\bm\omega}_M$ with ${\bm\omega}$ or ${\bm\Omega}$ or something else? This answer is far from obvious,"- he writes. Unfortunately, he forgot about one more frequency: the speed of rotation of the magnetic field ${\bm\omega}_H$. The magnetization ${\bf M}$ always follows
external magnetic field ${\bf H}$, so the rotation speeds of both vectors are the same, ${\bm\omega}_M={\bm\omega}_H$, and this is the only identity. Further he writes: ``The first magnetization relaxation equation accounting for the flow effect was proposed by Shliomis in ${\rm{1972}}$ \cite{Shliomis-72}:
\begin{eqnarray}
\frac{d{\bf M}}{dt}={\bm\omega}\times{\bf M}-\frac{\bf M-\bf M_0}{\tau_M}\,.
\end{eqnarray}
Here, [he continues] ${\tau_M}$ is called the Debye's magnetization relaxation time, which is often identified with [the Brownian one] $\tau_B$. Hence (?!), the Shliomis–Debye model identifies ${\bm\omega}_M$ with ${\bm\omega}$." This is a misunderstanding. As mentioned above, ${\bm\omega}_M$ coincides with the rotation frequency of the field, while the particle spin ${\bm\omega}$ satisfies equation
\begin{eqnarray*}
I_{MP}\frac{d{\bm\omega}}{dt}={\bf M}\times{\bf H} -6\eta\phi({\bm\omega}-{\bm\Omega}),
\end{eqnarray*}
which reduces -- due to the smallness of $I_{MP}$ -- to equality
\begin{eqnarray}
{\bm\omega}={\bm\Omega}+\frac{{\bf M}\times{\bf H}}{6\eta\phi}\,.
\end{eqnarray}
The latter makes it possible to exclude ${\bm\omega}$ from the equations of fluid motion and magnetization (9), the solution of which determines ${\bm\Omega}$ and ${\bf M}$, and thus also ${\bm\omega}$ by formula (10). In a constant field $H$, where always $\omega_M=0$, magnetic particles rotate in a vortex flow, so that $\omega\neq\omega_M$. The dependence ${\bm\omega}({\bm\Omega}, {\bf H})$ manifests itself in the magnetoviscosity, which for $\Omega\tau_B\ll 1$ is well described by the theory \cite{Shliomis-72}.

The purpose of including subsection IV~B "Primary dielectric peak of water" in the article is not clear. This subsection has nothing to do with hydrodynamics and/or ferrofluids and is completely devoted to the well-known and well-studied problem of molecular physics about the origin of the Debye absorption peak in water. The author is simply sharing his thoughts on the subject here.

Summarizing, an attempt to build ``a generic and consistent hydrodynamic theory of ferrofluids" based on relation (1) incompatible with the law of conservation of angular momentum was doomed to failure, which eventually happened.


\begin{thebibliography}{99}
\bibitem{notation}
The original transcription of Eq.~(59) in \cite{Fang-22} differs from Eq.~(7) because Fang used different notation: $L(\alpha)=m$ and $\alpha=L^{-1}(m)$.
\end{thebibliography}
\end{document}